\documentstyle[prl,twocolumn,aps,epsf,epsfig]{revtex}
\begin{document}
\draft


\title{Conductance as a Function of the Temperature in the  Double Exchange 
Model}

\author{M.J. Calder\'on$^{1,2}$, J.A.Verg\'es$^{1}$ and L. Brey$^{1}$}
\address{$^{1}$Instituto de Ciencia de Materiales de Madrid, 
Consejo Superior de Investigaciones Cient\'{\i}ficas.\\
28049 Cantoblanco, Madrid, Spain.\\
$^{2}$ Departamento de F\'{\i}sica Te\'orica de la Materia Condensada, 
Facultad de Ciencias, Universidad Aut\'onoma de Madrid, 
28049 Cantoblanco, Madrid, Spain.  }

\date{\today}

\maketitle

\begin{abstract}
We have used the Kubo formula to calculate 
the temperature dependence of 
the  electrical  conductance of the double exchange Hamiltonian. 
We average the conductance
over an statistical ensemble of clusters, which are  obtained by
performing Monte Carlo simulations on the classical spin orientation
of the double exchange Hamiltonian. 
We find that for electron concentrations bigger than 0.1, 
the system is metallic  at all temperatures.
In particular it is not observed any change in the temperature 
dependence of the resistivity near the magnetical critical temperature.
The calculated resistivity  near $T_c$ is around ten times
smaller than the experimental value. 
We conclude that the double exchange model is not able to explain the 
metal to insulator transition which experimentally occurs at temperatures near
the magnetic critical temperature.

\end{abstract}
PACS number 71.10.-w, 75.10.-b
\section{Introduction}
 
Materials that present extremely large magnetoresistance have potential
technological applications. As mixed valence compounds of the form 
$La ^{3+} _{1-x} A ^ {2+} _{x} Mn ^{3+} _{1-x} Mn ^{4+} _{x} O _{3} ^{2-}$
(where $A$ can be Ca, Sr or Ba) show  colossal magnetoresistance
there is a renewed interest in these oxides with perovskite
structure\cite{prim,rvw}.
There is a big correlation between the magnetic and the transport properties
of these oxide manganites. For $0.1\leq x \leq 0.5$ and low temperatures, 
the system is metallic and presents ferromagnetic order. 
As the temperature ($T$)
increases the system becomes insulator and paramagnetic. The metallic 
(insulator) behavior is defined in the sense that 
$d \rho / d T  > 0$ ($d \rho / d T  <0$), being  $\rho$ the electrical
resistivity. In the insulating phase the resistivity is bigger than the 
Mott resistivity $\sim 1000\mu \Omega  - cm$\cite{mott}.
The magnetic transition occurs at a $x$-dependent critical temperature
$T_c \sim 300K$.  The metal to insulator transition
occurs at a temperature very close to $T_c$.
For $x \rightarrow 0$ and low temperatures the system 
is a layer antiferromagnet with ferromagnetic coupling inside planes. At 
$x \leq 0.1$ phase separation between hole-rich
and hole-poor regions has been predicted\cite{yunoki2,dagotto,paco1}.
The electron-electron interaction plays an important role at low temperatures,
and it produces  charge ordering at particular values of $x <0.5$ and 
always for 
values of $x \geq 0.5$\cite{mori}.

In the $Mn$ oxides the electronically active orbitals are believed to be the
$Mn$ $d$ orbitals, and the mean $d$ occupancy is $4-x$. The $Mn$ ions are 
located at the corners of a simple  cubic lattice, and  they feel the
cubic crystal symmetry, which splits the $d$ orbitals into a $t_{2g}$ triplet
and a $e_g$ doublet. There is also a strong ferromagnetic 
Hund's rule coupling which align all electron spins in the $Mn$ $d$ orbitals.
The physical picture is that three electrons fill up
the $t_{2g}$  levels forming a core spin ${\bf S}$ of magnitude $3/2$
and the rest of the electrons go to the $e_g$ orbitals.
For small values of $x$, the perovskites show a long-range Jahn-Teller order
which selects a preferred combination of the $e_g$ orbitals and therefore it is
possible to assume that the electrons move only through one $d$ orbital.

To explain the ferromagnetism in these materials Zener\cite{zener}
introduced a double exchange (DE) mechanism, in which the electrons
get mobility between the $Mn$ ions using the magnetically inert oxygen
as an intermediate. The oxygens are located at the center of the lines
connecting the $Mn$ ions.
This conduction process is proportional to the electron transfer integral
and due to the strong ferromagnetic Hund's rule coupling 
it is maximum when the two cores spins involved in the process 
are parallel and it is zero
when they are antiparallel.
So in the DE model ferromagnetic coupling between $Mn ^{3+}$ and $Mn ^{4+}$
arises from the hopping of the electrons in the $e_g$ orbitals. Because
the alignment of spins on neighboring sites favors electronic motion, the
ferromagnetic ground state maximize the electron kinetic energy. 
When the temperature increases, the DE model undergoes a phase
transition towards a paramagnetic state. In this phase the core spins are
randomly oriented and fluctuate at frequencies related only to the
temperature. In the paramagnetic phase the electrons minimize 
their kinetic energy.
The DE model was more precisely formulated and extended by
Anderson and Hasegawa\cite{anderson} and by de Gennes\cite{degennes}.

Mean field theories have shown  that the ferro to paramagnetic
phase transitions at $T_c$ is accompanied by a change in the temperature
dependence of the resistivity\cite{kubo,furukawa,millis,millisp}.
However these calculations do not show metal to insulator transition 
at temperatures near $T_c$. These mean field calculations do not 
take into account the effect of the Berry phase arising from particle
motion in a spin background\cite{muller}.
Also the possible localization of electrons by the spin disorder in the
paramagnetic phase is neglected. Varma\cite{varma} proposed that random hopping
in the paramagnetic phase is sufficient to localize electrons and induce a 
metal to insulator transition at the magnetic critical temperature.
However Li {\it et al}\cite{li} have studied the mobility edge of the
DE model in the limit of $T \rightarrow \infty$ and they have found that 
random hopping alone is not enough to induce Anderson localization at the 
Fermi level in the range $0.1 \le x \le  0.4$.
Similar results were reported by Allub and Alascio\cite{alascio}
using the Ziman criterion\cite{ziman}. 

In this work we study the temperature dependence of the electric conductance 
of the DE model for different values of the electron concentration. 
We find that the system s metallic at all temperatures and, contrary
to the mean field results, we do not observe any feature in the
temperature dependence of the resistivity near the magnetic critical
temperature.

We calculate the dc conductance, $G$, by using the Kubo formula. 
We perform Monte Carlo
simulations on the classical spin orientations of the DE model\cite{calderon},
and for each electron concentration and temperature we obtain a
statistical ensemble of clusters of $Mn$ ions.
We obtain the conductance by averaging over several configurations. The
temperatures we are interested are $T \leq 500K$ ($T_c \sim 300K$) and these
temperatures are much smaller than the electron Fermi temperature
for $x \geq 0.1$. Therefore we always consider, both 
in the Monte Carlo simulations
and in the conductance calculations, that the electron temperature is zero.

In our model the $Mn$ ion spins are treated as classical and 
quantum
effects, as absorption and emission of spin density waves, 
are not correctly described. 
Although quantum effects should be important  at very low 
temperatures\cite{kubo}, 
they  do not matter  at temperatures near $T_c$.

The paper is organized as follows, in section II we introduce the 
DE Hamiltonian and describe the Monte Carlo calculations. In section III
we give some remarks about the use of the Kubo formula in the present context.
Section IV is dedicated to present and discuss the results of the paper, and
we
finish in section V with a summary.

\section{Double Exchange Hamiltonian.}

For large Jahn-Teller splittings, the electronic and magnetic properties of the
$Mn$ oxides are described by the following ferromagnetic Kondo
lattice Hamiltonian\cite{anderson},
\begin{eqnarray}
\widehat H = & - & t \sum _{<i,j>,\sigma} \left ( 
\widehat C ^ + _{i,\sigma} \widehat C _{j,\sigma}
+ h.c. \right  )  \nonumber \\
& - & J _H \sum _{i, \sigma , \sigma '}
\widehat C ^ + _{i,\sigma}\,  {\bf {\sigma}} _{ \sigma , \sigma '} \, 
\widehat C  _{i,\sigma '}
\cdot {\bf S} _i  \, \, \, \, ,
\end{eqnarray}
where $ \widehat C ^ + _{i,\sigma}$ 
creates an electron at site $i$ and spin $\sigma$,
$ {\bf S} _i$ represents the classical core spin at site $i$, $t$ is the
hopping amplitude between nearest-neighbor sites and $J_H$ is the
Hund's rule coupling energy. In the limit of infinite
$J_H$, Eq.(1) becomes the DE Hamiltonian,
\begin{equation}
\widehat H _{DE} = - \sum _{\langle i,j\rangle} \left 
( t _{i,j} \widehat C ^ {+} _i \widehat C _ j + h.c. \right )
\, \, \, \, \, ,
\end{equation}
Here $\widehat C^ {+} _i$ creates an electron at site $i$ with spin parallel 
to ${\bf S} _i$,
and the hopping amplitude acquires a Berry phase 
and it becomes a complex number given 
by\cite{muller},
\begin{equation}
t_{i,j}= t \left ( \cos {{\theta _ i } \over 2}
\cos {{\theta _ j } \over 2}
+\sin {{\theta _ i } \over 2}
\sin {{\theta _ j } \over 2}
e ^ { i ( \phi _i - \phi _j )} \right ) \, \, \, ,
\label{tcomplex}
\end{equation}
where $\theta _ i$ and $ \phi _i$ are the angles which characterize
the orientation of ${\bf S}_i$.
This complex hopping appears after rotating the conduction electron spins
so that the spin quantization axis at site $i$
is parallel to  ${\bf S} _i$, and then project onto the spin parallel
to ${\bf S} _j$. 

In order to calculate the electrical conductance, for each $T$ and $x$, it
is necessary a statistical ensemble of clusters, which simulates 
the thermal fluctuations. Each cluster is characterized by a set of core
spin $\{ {\bf S} _i \}$ and its chemical potential $\mu$.
We obtain the clusters by performing Monte Carlo simulations on the 
variables $\theta _i$ and $\phi _i$ of the Hamiltonian Eq.(2).
In each Monte Carlo step it is necessary to diagonalize a matrix of size
equal to the number of $Mn$ ions in the clusters and the DE energy is the sum 
of the eigenenergies of the occupied electron levels. In this process we assume
that the electron Fermi energy is much bigger than the temperatures of interest
in this work
(at $x \sim$0.2, $T_{Fermi} \sim$ 2000$K$ and
$T_c \sim$ 300$K$).
The diagonalization imposes a restriction on the dimension
of the unit cell used in the simulations.
Recently two of us have studied, by using Monte Carlo simulations,  
the magnetic phase diagram of the
DE model\cite{calderon}.
To avoid finite size problems in the simulation, the following 
approximation for the energy has been obtained and used for the DE
energy,
\begin{equation}
 E  \simeq 
-2 t \langle \widehat C ^ + _i \widehat C _j 
\rangle _0  \sum _{\langle i,j\rangle} \cos { { \theta _{ij}} \over 2} -
 a_2 
 \sum _ { \langle i,j\rangle} 
(\bar {t} - t _ {i,j}) ^ 2 \, \, \, .
\label{energy-approx}
\end{equation}
Here $\langle \widehat C ^ + _i \widehat C _j \rangle _0$ 
and $ a_2$ are quantities which
depend on $x$ and 
do not depend on temperature,  $\bar {t}$ is the average of the
absolute value of the hopping amplitude and $\theta _{i,j}$ represents
the angle formed by the core spins located at sites $i$ and $j$.
In reference \cite{calderon} it was showed that expression (4) 
is a good approximation
to the DE kinetic energy.
In that work it was also concluded that 
the magnetic critical temperatures of the
DE model are in the range of the experimental ones, and  it   was  
obtained that the complex phase of the hopping amplitude has a negligible effect
in the value of $T_c$.

Using the conclusions of the cited work, in this paper we 
perform  Monte Carlo simulations  on the variables 
$\theta _i$
and $\phi _i$, using the expression (4) for the DE energy.
The simulations are performed in $N \times N \times N$ cubic lattices.
Technical details about the Monte Carlo calculations are given in reference
\cite{calderon}, here just  say that typically 5000-7000 Monte Carlo
steps per spin are used for thermal equilibration. 
The different clusters
which form the statistical ensemble are chosen every 100 steps per spin after
equilibration.

\section{D.c. conductance via Kubo formula}
In calculating the conductance of the system we use the standard Kubo
formula \cite{nozieres,datta}. The static electrical conductivity 
at chemical potential $\mu$ is given by
\begin{equation}
G = \sigma_{zz}(0)=-2 \frac{e^2}{h} {\rm Tr} 
\left [
(\hbar {\hat v_z}) 
{\rm Im}\, 
{\mathcal \widehat G}
(\mu )
(\hbar {\hat v_z}) {\rm Im}\, 
{\mathcal \widehat G}
(\mu )
\right ]
\; ,
\label{kuboe}
\end{equation}
\noindent where ${\rm Im\,}{\mathcal \widehat G}(\mu )$ is calculated from 
the advanced and retarded Green functions
$$
{\rm Im\,}{\mathcal \widehat G}(\mu  )=\frac{1}{2i}\left[{\mathcal 
\widehat G}^{R}(\mu  )-{\mathcal \widehat G}^{A}(\mu   )\right ] \; ,
$$
and the velocity (current) operator ${\hat v_z}$ is related 
to the position operator ${\hat z}$ through the equation of motion
\begin{equation}
\hbar {\hat v}_z = \left [ {\widehat H},{\hat z} \right ] \; ,
\label{motion}
\end{equation}
$\widehat{H}$ being the Hamiltonian in Eq. (2).

Numerical calculations are carried out for a bar geometry
connecting the $N \times N  \times N $ cluster 
to two ideal semiinfinite leads of $N  \times N  $
cross section. This provides complex selfenergies at opposite
sides of the sample \cite{datta}. They are first calculated
for the normal modes of the lead and then transformed
to the local tight-binding basis. Specifically,
the retarded selfenergy due to the mode of wavevector
$(k_x, k_y)$ at energy $\varepsilon $ is given by:
$$
\Sigma(k_x,k_y)={1 \over {2 t}}
\left( \varepsilon -\varepsilon (k_x, k_y)
-i\sqrt{4t^2-(\varepsilon - \varepsilon (k_x, k_y))^2} \right) \; ,
$$
within its band and by:
$$
\Sigma(k_x,k_y)={1 \over {2 t}}
\left( \varepsilon - \varepsilon (k_x, k_y) \mp 
\sqrt{(\varepsilon -\varepsilon (k_x, k_y))^2-4t^2} \right) \; ,
$$
outside the band (minus sign for $\varepsilon > \varepsilon (k_x, k_y)$
and plus sign for $\varepsilon < \varepsilon (k_x, k_y)$), where
$\varepsilon (k_x, k_y)=2 t(cos (k_x)+ cos (k_y))$ is the
eigenenergy of the $(k_x, k_y)$ mode).
The transformation from normal modes to the local tight-binding basis
is obtained from the amplitudes of the normal modes:
$$
<(n_x, n_y)|(k_x, k_y)>={2 \over {N +1}}
\sin(k_x n_x) \sin(k_y n_y) \; ,
$$
where $n_x$ and $n_y$ represent the tight-binding orbital position.
Once the selfenergy matrices introduced by the left (right)
semiinfinite leads are determined,
the retarded Green function matrix  of the sample 
${\mathbf G}(\varepsilon )$
is defined by the following set of
$N  \times N   \times N $ linear equations:
\begin{equation}
[\varepsilon  {\bf I} - {\bf H} - {\bf \Sigma}_{\mathrm l}
(i\varepsilon ) - {\bf
\Sigma}_{\mathrm r}(\varepsilon )] {\mathbf G}(\varepsilon ) = {\bf I} ~~~,
\label{green1}
\end{equation}
where ${\bf \Sigma}_{\mathrm l(r)}(E)$ stand for the
selfenergy matrices introduced by the left (right) semiinfinite leads.
This set of equations is efficiently solved using a layer by layer
inversion scheme that takes advantage of the band structure of
the coefficients matrix. The advanced Green function matrix
is simply the conjugate of the transpose of the retarded one.

The last ingredient that is necessary for the evaluation of the Kubo
formula is the velocity operator. It is obtained through Eq.(\ref{motion})
once the position operator $\hat z$ is known. This operator is
determined by the spatial shape of the electric potential energy.
Taking advantage from the fact that
the detailed form of the electric field does not matter
within one-electron linear response theory,
an abrupt potential drop at one of the two cluster 
sides provides the simplest
numerical implementation of the Kubo formula \cite{verges}.
Certainly, Eq.(\ref{motion}) shows that nonvanishing elements
of the velocity operator are restricted to the 
two  layers at the sides of the potential drop. Furthermore,
the trace appearing in Eq.(\ref{kuboe}) makes the knowledge of
the Green functions on the same restricted set of sites
enough for the evaluation of the conductance. Consequently,
Green functions are just evaluated for two
consecutive layers at the cluster boundary.

\section{Results}

We start showing the results corresponding to a perfect ferromagnetic system. 
In Fig.1a we plot the conductance as a  function of the chemical potential
for a cluster of size $N$=20 at $T$=0. The conductance is finite for
chemical potentials in the range  $-6t \leq \mu \leq 6t$. This is the range
of energies where the density of states of the perfect system is finite, see
Fig.1b. Since at $T$=0 all core spins are aligned, the transport in the system
is ballistic and the conductance is just limited by the size of the cluster.
Therefore, at $T$=0 the conductance is proportional to the number of 
transport channels at the Fermi energy, which for large clusters increases
as $N^2$.
We have verified this behavior for $N \geq 10$ and also we have checked
that for  system sizes bigger than $N$=10, the overall shape of $G$ 
does not depend on the cluster size.

In the opposite limit we have calculated the conductance in the 
paramagnetic phase at $T \rightarrow \infty$. In this limit the 
orientation of the local spins $\{ {\bf S} _ i \}$ is random. We choose
${\bf S} _i$ to be uniformly distributed on a sphere, i.e. the probability
$P$ of having a core spin with azimuthal angle $\phi _i$ is $1/2\pi$, whereas
the polar angle distribution verifies $P(\cos {\theta _i})$=1/2.

\begin{figure}
\epsfig{figure=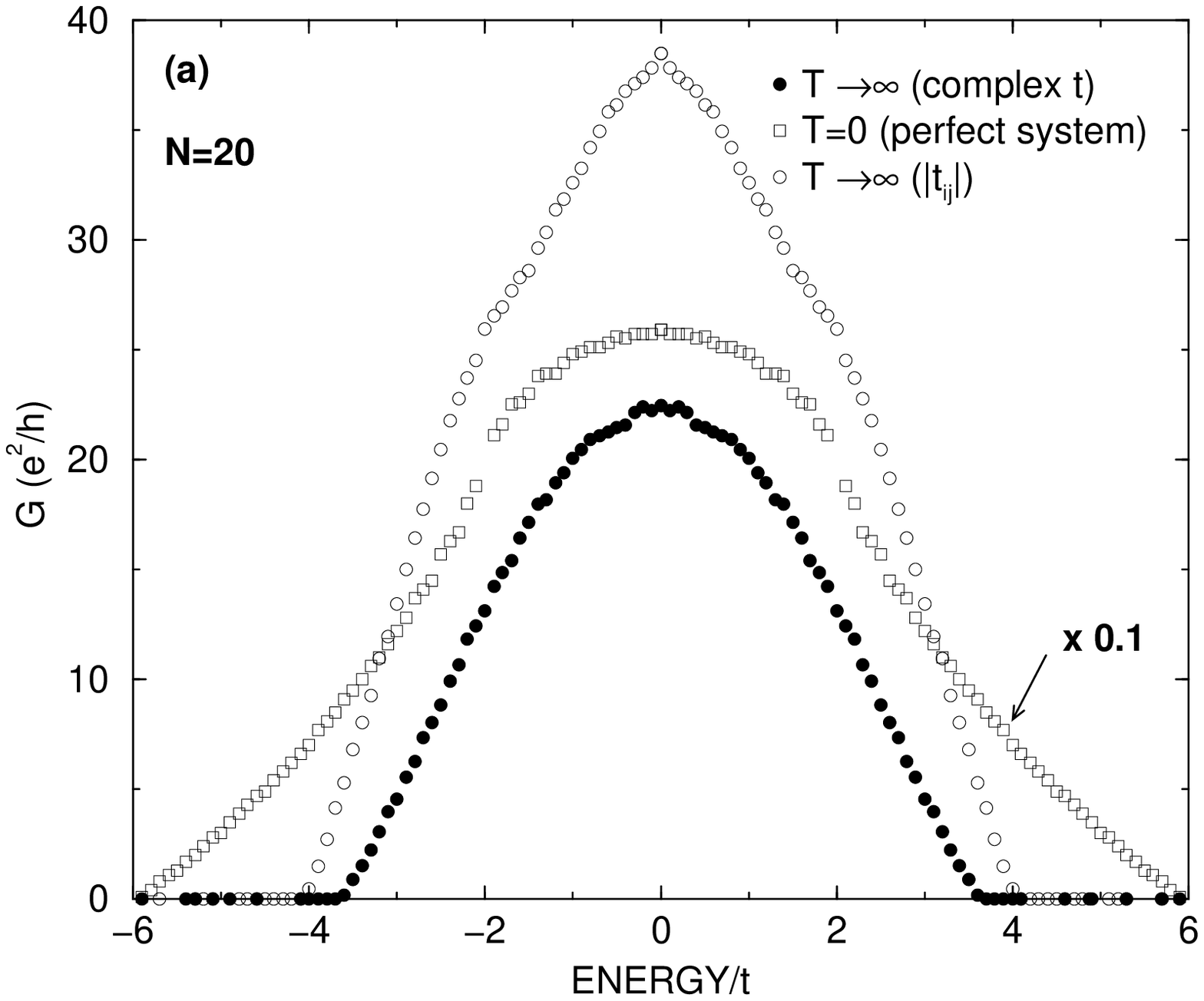,width=8.5cm}
\epsfig{figure=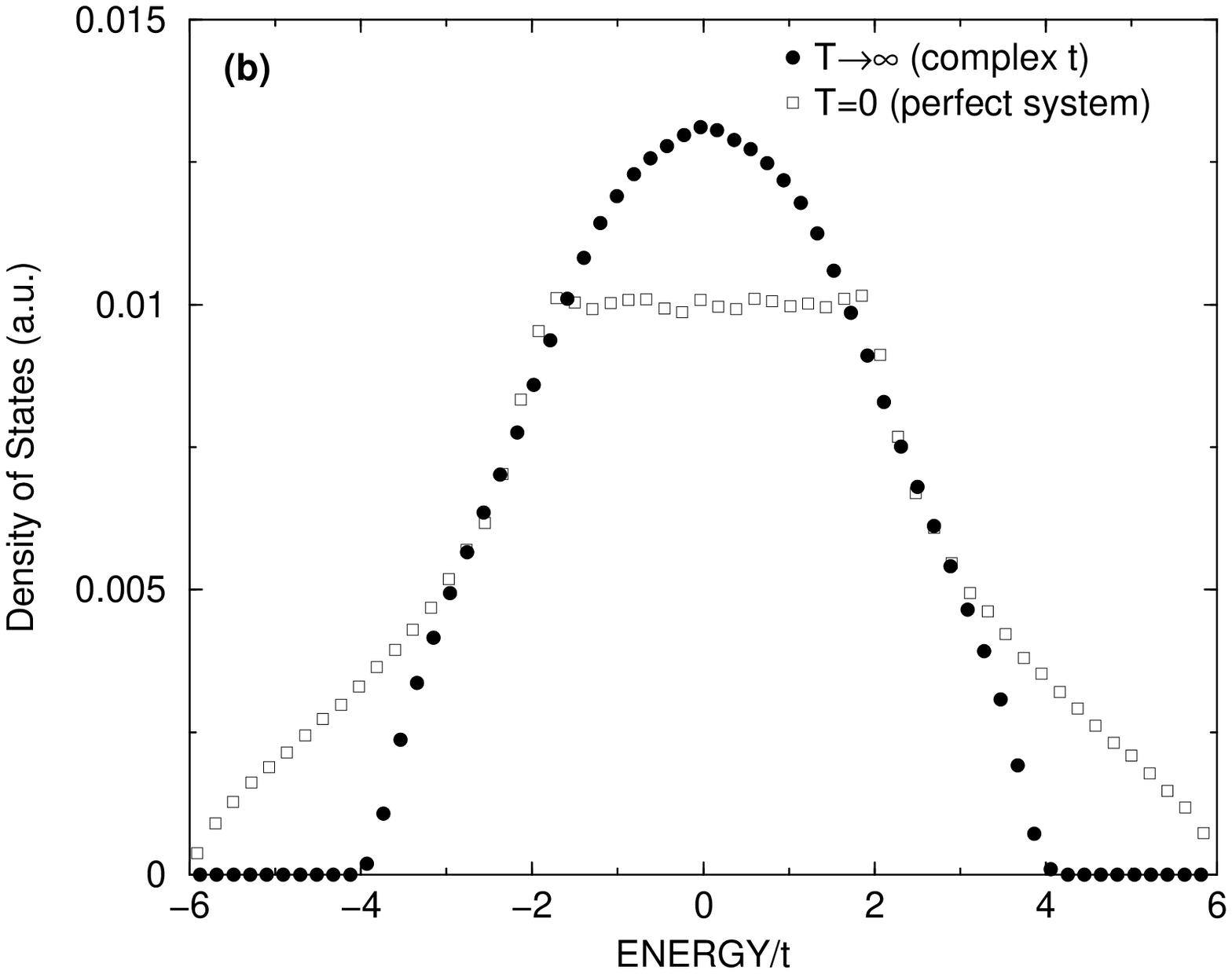,width=8.5cm}
\caption{
a) Energy dependence of the conductance for the $T=0$ case (empty squares),
the $T \rightarrow \infty $ case (filled dots), 
and the $T \rightarrow \infty $ case 
without Berry's phase (empty dots).
b) Density 
f states as a function of the energy for the $T=0$ case (empty squares)
and the $T \rightarrow \infty $ case (filled dots).} 
\end{figure}

In Fig.1 we plot  the 
conductance and 
the density of states as a function of the energy in the 
$T \rightarrow \infty$ limit. 
This quantities are averaged on many configurations of clusters of size
$N$=20. The density of states shows  an effective band edge at 
$E_b \simeq - 4 t$. This occurs because the average value of the absolute value
of the hopping amplitude at $T\rightarrow \infty$ is $<|t_{i,j}|>$=2/3$\, t$. 
Although the band edge starts at $-4t$, $G$ is different from zero only for 
values of the chemical potential in the range $-3.6 t\leq \mu \leq 3.6 t$. 
The difference between the band edge energy and the minimum energy with 
$G \neq 0$ is due to the fact  that all the states with energy $-4t <E <-3.6t$
are localized and do not contribute to $G$. Therefore $E_c \simeq $-3.6$t$ is the 
$T\rightarrow \infty $ mobility edge of the DE model. This result is in agreement with 
localization length calculations\cite{li}, which located the mobility edge
at $ |E_b |\simeq 3.56 t$. 
Our calculations show that $|E_b|$
is constant for values of $N$ bigger than $N$=10. This 
implies that 
for energies lower than $E_b$ 
the localization length  
is  
shorter than 10 lattice parameters. 
For  $N \geq$10, the overall shape of the 
average conductance as a function of energy
is almost independent of the cluster size, 
and its value increases linearly with $N$.
This implies that in the $T \rightarrow \infty$ limit the system is metallic, 
the electron transport is diffusive and the system verifies the
Ohm's law: the cluster resistance is proportional
to the cluster length and inversally proportional to the
cross-sectional area. The proportionality constant is the 
resistivity $\rho$. 

\begin{figure}
\epsfig{figure=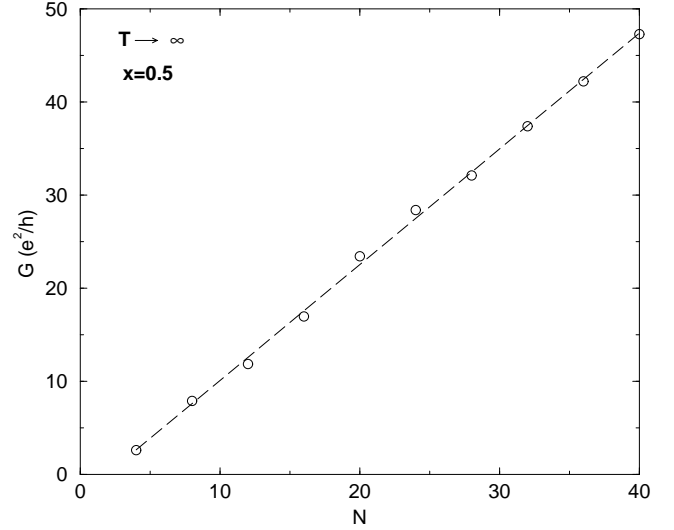,width=8.5cm}
\caption{
Conductance as a function of the cluster size, for $x=0.5$ and $T\rightarrow \infty$.
The dashed line is only a guide to the eye.}
\end{figure}

As an example of this behavior we plot in Fig.2
the average conductance at $\mu=0$ as a function of the cluster size
$N$. The linear behavior of $G$ versus $N$ down to $N$=4, implies
that the elastic mean free path is smaller than 4 lattice parameters.
The slope of the straight line is the conductivity, in lattice
parameter units, of the DE model at $T\rightarrow \infty$ and $x$=0.5.

In Fig.1 we also plot the average conductance 
in the $T \rightarrow \infty $ limit of the DE model but neglecting the Berry's
phase in the hopping amplitude i.e. $t_{i,j} \rightarrow |t_{i,j} |$. In this case
$G$ is bigger than in the case of complex hopping.
Also it seems that the mobility edge appears at lower energies than in 
the complex hopping DE model.

The previous calculations correspond to the $T$=0 and 
$T \rightarrow \infty$ limits. Now we present the results for $G$ as a
function of $T$, for different values of $x$.
In these case the average of $G$ is done with an statistical ensemble
of clusters which are 
obtained from Monte Carlo calculations as described in Section II.
When $T$ increases the disorder in the system increases and two
main effects occurs: a) a $T$-dependent mobility edge appears at 
low energies and b) the extended states acquire an elastic 
mean free path, $\ell$,
which decreases with $T$.

\begin{figure}
\epsfig{figure=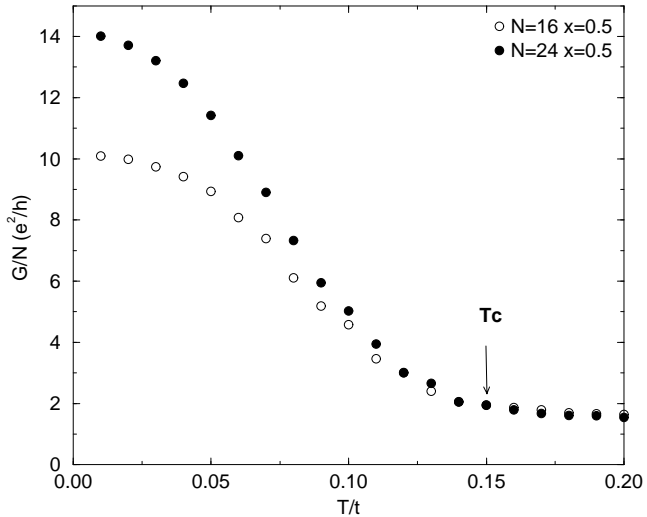,width=8.5cm}
\caption{
Temperature dependence of the conductivity for the case of $x=0.5$ and 
$N=16$ and $N=24$. 
The magnetic critical temperature $T_c$ is pointed with an arrow. 
Above this temperature conductivity is independent of $N$.}
\end{figure}
\begin{figure}
\epsfig{figure=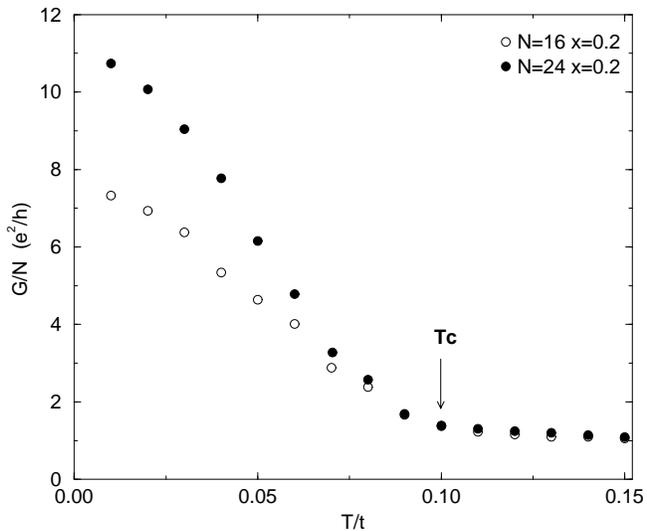,width=8.5cm}
\caption{
Temperature dependence of the conductivity for the case of $x=0.2$ and 
$N=16$ and $N=24$. 
}
\end{figure}
With respect to point a) we know  that in the maximum disorder case ($T\rightarrow
\infty$), the mobility edge is $|E_c| \sim 3.6t$. This energy is very close
to the effective band edge and 
only less than 0.5$\%$ of the total states are localized.
For lower temperatures we expect  $E_c$ to  be closer to the 
effective band edge
and the number of localized states should be even smaller. We are interested in 
values of $x>$0.1 for which the chemical potential is much higher than
$E_c$ and 
for these electron concentrations we expect the system to be  metallic at any $T$.
The DE model is an insulator only at very small values of the
electron concentration, $x<$0.05 and high temperatures.
With respect to the elastic 
mean free path, point b), we find that the conductivity of the
system decreases continuously with $T$, until it reaches the  
$T \rightarrow \infty$ limit (note that $G \sim \ell$).
On the contrary if $\ell$ is bigger than the cluster size the transport is  ballistic,
the conductance is 
determined by the cluster size, and increases as $N^2$. 
When $\ell$ is shorter than $N$ the transport is diffusive and the conductance
is proportional to $N$.

In Fig.3 we plot the conductance divided by $N$ for $x$=0.5 and two values of the
cluster size, $N$=16 and $N$=24. For low temperatures the core spins disorder is very
weak and the elastic mean free path is larger than the cluster size. In this regime
the transport  is 
ballistic and the conductivity increases linearly with $N$.
For large values of the temperature, $G/N$ is 
independent of $N$. This is because
the elastic mean free path is shorter than the cluster size and $G/N$ is the electrical
conductivity 
in units of the system lattice parameter.
For $T>0.12t$ the values of $G/N$ obtained using $N=16$ and $N=24$ coincide, this 
implies that for these temperatures $G/N$ is the conductivity.
For $x=0.5$ the magnetic critical temperature is $T _c \sim 0.15t$, and we can see in Fig.3 that the conductivity is a smooth function of $T$, practically constant, near $T_c$.

In Fig.4 we present the same quantity $G/N$ as a function of $T$ for $x =0.2$. 
Also for this electron concentration the conductivity near $T_c$ is a smooth 
function of $T$  and the system is metallic.
We have obtained similar results for different values of $x>0.1$.
The values for $T_c$ are taken from reference\cite{calderon}.

\begin{figure}
\epsfig{figure=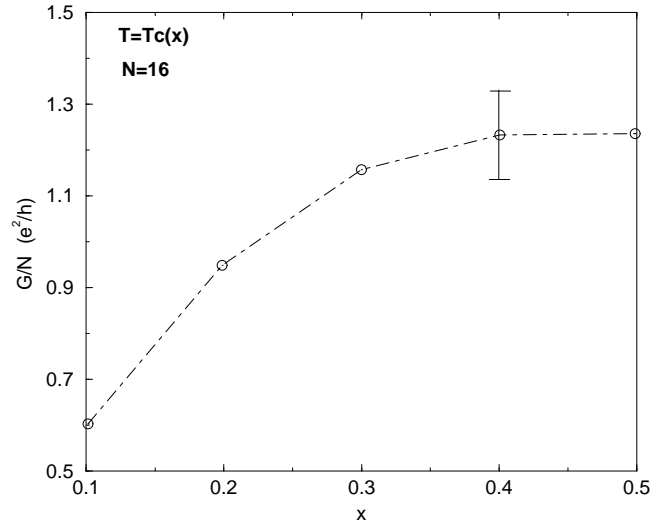,width=8.5cm}
\caption{
Conductivity, evaluated at $T_c$, as a function of $x$. 
The dashed line is only a guide to the eye.}
\end{figure}
\newpage

In Fig.5 we plot as a function of $x$, the value of the conductivity in lattice units 
evaluated at $T_c$. 
Using the value of $4 \AA$ for the lattice parameter, 
we obtain a resistivity $\rho \sim 0.001 \Omega - cm$
at $x=0.2$ near $T_c$. This resistivity is around ten times smaller than the 
experimental ones, which also have an insulator-like dependence with $T$.
This is a clear indication that it is necessary to add other terms to the
double exchange Hamiltonian, in order to explain the occurrence of the
metal insulator transition at temperatures near the magnetic critical 
temperature\cite{millis1}.

\section{Summary}
We  have calculated the temperature dependence of 
the  electrical  conductance of the double exchange Hamiltonian. 
The Kubo formula has been use for the calculation of
the conductance. Conductance is defined as  the average conductance  
over an statistical ensemble of clusters. These clusters are obtained by
performing Monte Carlo simulations on the classical spin orientation
of the double exchange Hamiltonian. 
The calculations have been done for different electron concentrations.
We find that the system is metallic at all temperatures and, contrary to 
the mean field calculations, we do not observe any change in the temperature 
dependence of the resistivity near the magnetical critical temperature.
Near $T_c$ the resistivity we obtain is around ten times
smaller than the experimental value. 
We conclude that the double exchange model is not able to explain the 
metal to insulator transition which experimentally occurs at temperatures
near the magnetic critical temperature. Other effects not included in the DE 
Hamiltonian, as electron-electron interaction or electron- phonon interaction, 
are needed in order to understand the electrical behavior of the
oxide $Mn$.

{\it Acknowledgments.}
We thank Luis Mart\'{\i}n-Moreno for useful discussions.
This work was supported by the CICyT of Spain under Contract No. PB96-0085.
MJC and LB also acknowledge financial support from the
Fundaci\'on Ram\'on Areces.



%

\end{document}